\documentstyle[12pt]{article} 
\begin{document}
\voffset -0.4 true cm
\hoffset 1.5 true cm
\topmargin 0.0in
\evensidemargin 0.0in
\oddsidemargin 0.0in
\textheight 8.7in
\textwidth 7.0in
\parskip 10 pt

\def\half{{1 \over 2}}

\newcommand{\eeas}{\end{eqnarray*}}
\def\laa{\langle\kern-.3em \langle}
\def\raa{\rangle\kern-.3em \rangle}
\def\la{\lambda}
\def\cn{{\cal N}}\def\cz{{\cal Z}}
\def\ymt{Yang-Mills theory}
\def\sftH{string field theory Hamiltonian}
\def\aaa#1{a^{\dagger}_{#1}}
\def\aap#1#2{(a^{\dagger}_{#1})^{#2}}
\def\ee{\hbox{e}}
\def\dd{\hbox{d}}
\def\tr{\hbox{tr}}
\def\Tr{\hbox{Tr}}
\def\part{\partial}
\def\dj#1{{\delta\over{\delta J^{#1}}}}
\def\cO{{\cal O}}
\def\ssc{\scriptscriptstyle}
\begin{titlepage}

\begin{flushright}
PUPT-1924\\
hep-th/0003179 \\
\end{flushright}

\vskip 1.2 true cm

\begin{center}
{\Large \bf Schwinger-Dyson = Wheeler-DeWitt:\\
gauge theory observables  as bulk operators}
\end{center}
\vskip 0.6 cm

\begin{center}

Gilad Lifschytz and Vipul Periwal

\vspace{3mm}

{\small \sl Department of Physics} \\
{\small \sl Joseph Henry Laboratories} \\
{\small \sl Princeton University} \\
{\small \sl Princeton, NJ 08544, U.S.A.} \\
\smallskip
{\small \tt gilad@viper.princeton.edu, vipul@princeton.edu}

\end{center}

\vskip 0.8 cm

\begin{abstract}

\noindent

We argue that the second-order gauge-invariant Schwinger-Dyson
operator of a gauge theory is
the Wheeler-DeWitt operator in the dual string theory. Using
this identification,  we construct a set of
operators in the gauge theory that correspond to excitations of
gravity in the bulk.
We show that these gauge theory operators have the expected properties
for describing the semiclassical local gravity theory.
\end{abstract}

\end{titlepage}

\section{Introduction}

Following Maldacena's conjecture\cite{malda,gkp,wmalda}, 
a  fair amount of evidence has
accumulated in support of the identification of certain
(supersymmetric) gauge theories
in the large $N$ limit with (super)gravity theories on appropriate backgrounds
with one additional non-compact direction.
The existence of a semiclassical (super)gravity
background is predicated on taking the 't Hooft coupling
$g^{2}_{\scriptscriptstyle \rm YM}N\equiv \lambda$ to
be very large, which makes any direct quantitative
verification of the conjecture difficult.

If large $N$, large $\lambda$ gauge theories are dual to classical
gravity backgrounds, we should be able to construct
bulk operators in the gravity theory, for example to study the
properties of singularities.  However, correlation functions of the gauge
theory are
boundary correlations in the gravity theory, and these are believed to
be the only observables.  The obvious question that needs to be
addressed then is: How is bulk information encoded in boundary correlation
functions?  In fact we do not yet understand well enough even 
the relationship between 
classical gravity concepts and the dual gauge theory, although 
some understanding
on issues like causality and horizons have been gained\cite{caubh}.
While the identification
of the extra large dimension with some energy scale\cite{malda}\ 
has enabled one to
deduce certain properties of objects in the bulk  as seen
from the gauge theory, a map between local bulk operators and gauge theory 
observables has not been found. The case of 
a classical non-interacting field on the AdS background was discussed 
in \cite{bdhm}.
 
This is intimately related to an intriguing aspect of AdS/CFT duality,
the idea of holography, first proposed by 't Hooft\cite{th}.
Briefly, black hole thermodynamics  suggests that the
degrees of freedom inside the horizon of a black hole reside at the
horizon.  This is difficult to reconcile with the extensivity that
we expect in garden-variety low-energy effective field theories.  
't~Hooft and Susskind\cite{th}\
suggested that this behaviour might be a general feature
of quantum gravity.  In fact, the AdS/CFT duality turns out to
be an explicit example of holography\cite{wmalda,witsus}: 
The (super)gravity  theory
is defined  in five noncompact dimensions for four-dimensional gauge theory.

Some time ago, Ishibashi, Kawai and their collaborators\cite{ik}\ 
constructed toy models of quantum gravity coupled to $c<1$ matter
in temporal gauge.  It turned out that the string field theories they
constructed were related 
to matrix models\cite{jevic}\ via the stochastic quantization of Parisi and 
Wu\cite{pw}.  Extrapolating 
these ideas to gauge theories using the observation of 
Marchesini\cite{marc}\ led to a proposal for a direct nonperturbative 
connection 
between gauge theories and string theories in temporal gauge\cite{v1},
making contact with Polyakov's ideas regarding gauge theories and 
noncritical strings\cite{pol}.

We consider in this paper the proposition that the second-order
gauge-invariant Schwinger-Dyson equations of the gauge theory 
are the Wheeler-DeWitt equations of string theory, as Ishibashi {\it 
et al.}\cite{ik}\ did for the $c<1$ models.  
We explain why such an identification
depends crucially on  the structure of
the Schwinger-Dyson equations expressed in terms of Wilson loops.
This identification allows us to
construct gauge theory observables that are naturally related to
operators in the bulk of the dual gravity background, shedding
some light on the manner in which holography is implemented in the
AdS/CFT duality.  We
show that these operators have the expected properties, using the fact
that the second-order Schwinger-Dyson equations are in fact just the
equilibrium conditions of stochastic quantization, with the operator
that generates the Schwinger-Dyson equations also the Fokker-Planck
Hamiltonian of the gauge theory. This connection has already been used
to suggest a connection between the radial direction of AdS and
stochastic time, and to explain the finite $N$ truncation of the spectrum of
chiral primary operators in the gauge theory\cite{lp}\
from the string theory perspective.  Some related work can be found in
\cite{others}.

The plan of this paper is as follows: We start with a brief account of
the relevant results in the stochastic quantization of gauge theories.
We then motivate an identification of the
Schwinger-Dyson operator of the gauge theory
with the Wheeler-DeWitt operator of the gravity theory. Using this
identification a set
of ``bulk'' operators is constructed from the gauge theory and their
properties explored.

\section{Stochastic Quantization}

Stochastic quantization\cite{pw,stoc}\ is a representation of 
quantum correlation
functions as equilibrium values of correlation functions of
a classical system coupled to white noise.

Given a classical equation of motion for an Euclidean field theory 
in $d$ dimensions,
we associate a Langevin equation
\begin{equation}
 \frac{\partial\phi}{\partial \tau}=-\Omega \frac{\delta S}{\delta \phi}
+ \eta(x,\tau).
\label{lan}
\end{equation}
where $x$ is a point in $d$ dimensions,
$\tau$ is a fiducial Euclidean `time' coordinate,
$\Omega $ is a time scale, and
$\eta$ is white noise with a noise ensemble average
\begin{equation}
	<\eta(x,\tau)\eta(x',\tau')>_{\eta}  =
\Omega \delta(x-x')\delta(\tau-\tau').
\end{equation}
(We will set $\Omega=1$ for most of the paper.) 
The basic statement of stochastic quantization is that
the equal $\tau$ equilibrium  stochastic correlation
functions are equal to the
quantum correlation functions of the original theory:
If $\phi_{\eta}$ is the solution of the Langevin equation (\ref{lan}) 
with some initial condition 
then
\begin{equation}
	\lim_{\tau\uparrow\infty} <\prod_{i}\phi_{\eta}(x_{i},\tau) >_{\eta}
=\langle \prod_{i}\phi(x_{i})\rangle,
\end{equation}
where the left hand side is the stochastic average, which is 
independent of the initial condition in the large $\tau$ limit,
and the right hand side
is the vacuum correlation function in the quantum field theory.
Equal $\tau$ equilibrium ($\tau\uparrow\infty$)
correlation functions are, of course, just a particular
case of more general unequal $\tau$ correlation functions:
\begin{equation}
< \phi_{\eta}(x_1,\tau_1)\ldots \phi_{\eta}(x_n,\tau_n)>_{\eta}\equiv
\int d \eta(x,t) e^{-\int\eta^2 /\Omega} \phi_{\eta}(x_1,\tau_1)\ldots
\phi_{\eta}(x_n,\tau_n).
\end{equation}

An equivalent formulation of stochastic quantization is obtained by
finding  a stochastic action $S_{\scriptstyle{\rm 
stoc}}[\phi(x,\tau)]$ such that the
correlation functions computed in a functional integral with this
action
are the stochastic correlation functions. For a scalar field theory
\begin{equation}
S_{\scriptstyle \rm
stoc}=\int \dd\tau \dd x \left[{1\over2}\left(\frac{\dd\phi}{\dd\tau}\right)^2+
{1\over 8}\left(\frac{\delta S}{\delta
\phi}\right)^2-{1\over 4} 
\left(\frac{\delta^{2} S}{\delta\phi(x)^{2}}\right)\right] .
\end{equation}
The last term is divergent and must be understood in the context of 
a regularization.  In dimensional regularization, it is set to zero as 
usual, but this choice of regularization may not be appropriate in 
all instances.
Associated with this action is a Hamiltonian
which generates translations in the $\tau$ direction.
This is the Fokker-Planck (FP) Hamiltonian
\begin{equation}
H_{\ssc\rm FP} = \int \dd x \left[{\delta \over\delta\phi(x)}-{\delta
S\over\delta\phi(x)}\right]{\delta \over\delta\phi(x)}.
\end{equation}
This Hamiltonian is not hermitian but becomes hermitian under a
similarity
transformation by $\ee^{-S/2}$ 
\begin{equation}
\hat{H}_{\ssc\rm FP}=\int \dd x \left[-\frac{1}{2}\frac{\delta^2}
{\delta\phi^{2}} +
\frac{1}{8}\left(\frac{\delta S}{\delta \phi}\right)^2 -\frac{1}{4}
\frac{\delta^2 S}{\delta\phi^{2}}\right].
\end{equation}
Introducing a source $J$ for $\phi,$  we can define
a second-order Schwinger-Dyson (SD) operator $H_{\ssc\rm SD}$ by
\begin{equation}
\langle H_{\ssc\rm FP} \ee^{\int J \phi}\rangle \equiv H_{\ssc\rm
SD}\left(J,\frac{\delta}{\delta J}\right)
\langle \ee^{\int J \phi}\rangle \ ,
\label{fpsd}
\end{equation}
where   $\langle \ldots \rangle$ denotes the expectation
value in the field theory.
The equal $\tau$ correlation functions can also be represented 
as (we have set $\Omega=1$)
\begin{equation}
< \prod_{i}\phi_{\eta}(x_{i},\tau) >_{\eta} = \laa
0|\ee^{-\tau H_{\rm FP}} \phi(x_1) \ldots \phi(x_n)|0\raa,
\label{cor1}
\end{equation}
where in the large $\tau$ limit one gets the correlation functions of
the original field theory, and we can use either $H_{\rm FP}$ or
$\hat{H}_{\ssc\rm FP}$.
The existence  of the large $\tau$ limit implies
\begin{equation}
\laa 0|\ee^{-\tau H_{\ssc\rm FP}}  H_{\ssc\rm FP}
\phi(x_1) \ldots \phi(x_n)|0\raa =0;
\label{equil}
\end{equation}
these are the equilibrium conditions of stochastic quantization.
The expectation value in equation~(\ref{cor1}) is defined with respect
to a formal vacuum state (which depends on whether one  uses
$H_{\ssc\rm FP}$ or $\hat{H}_{\ssc\rm FP}$) satisfying
\begin{eqnarray}
\laa 0|H_{\ssc\rm FP}^{\dagger} =0\\
H_{\ssc\rm FP} |0\raa  =0
\end{eqnarray}

We shall be interested in another set of correlation functions,
those in which the stochastic time is taken to infinity but the correlation
functions depend on finite differences of stochastic time between the
operators:
\begin{equation}
\lim_{\tau\rightarrow\infty}<{\cal O}_{1}(x_1,\tau+t_1){\cal
O}_{2}(x_2,\tau+t_2)\ldots {\cal O}_{n}(x_n,\tau)>_{\eta } .
\label{scf}
\end{equation}
While finite $\tau$ correlation functions cannot be directly interpreted as
quantum correlations, the correlation functions in 
equation~(\ref{scf}) 
are in fact
quantum correlations  in the original theory. They contain no  new
information
of course since they can be written
as a sum of equal time correlation functions with coefficients
depending on $t_{i}-t_{j}.$  These correlations are translation
invariant in $t_{i}.$

Turning to  gauge theories, the
Fokker-Planck hamiltonian for gauge theories
\begin{equation}
	\int\dd x \ {1\over N}
\sum_{\mu,a}\left({\delta\over{\delta A_{\mu}^a(x)}} - {{\delta S}\over{\delta
A_{\mu}^a(x)}}\right){\delta\over{\delta A^{\mu a}(x)}}
\label{gaugesd}
\end{equation}
has the property that it is gauge-invariant.  Acting on Wilson loops,
the action of this operator (or of its associated Schwinger-Dyson
operator) can be interpreted as Wilson loop
diffusion, joining and splitting processes.

A physical interpretation of stochastic time is obtained by
introducing Schwinger proper-time representations of propagators.
Viewing all Feynman diagrams as starting from external legs at
$\tau=0,$ the time evolution is generated by the Fokker-Planck
Hamiltonian\cite{ikspecial}.  In this interpretation, 
a finite $\tau$ amplitude is a
sum over Feynman diagrams with the restriction that the total
proper-time is bounded.  Thus finite $\tau$ amplitudes are infrared
regulated.  For example,
for a free massless scalar field the two-point function at finite
stochastic time
(with appropriate boundary conditions) is
\begin{equation}
<\phi_{\eta}(k,\tau) \phi_{\eta}(-k,\tau)>_{\eta}=\frac{1}{k^2}(1-\ee^{-k^2
\tau})=\int_{0}^{\tau} \dd s\ \ee^{-sk^2}
\label{finiteir}
\end{equation}
which exhibits no singularity as $k^{2}\downarrow 0.$

\subsection{The Schwinger-Dyson equations}

\def\dj#1{{\delta\over{\delta J^{#1}}}}
\def\cO{{\cal O}}
In gauge theories, there is a natural complete set of gauge invariant
operators. These are
Wilson lines $\cO_{C}\equiv\Tr P\ee^{i\oint_{C} A}$ with $\Tr$ a normalized
trace defined by $\Tr (1) = 1.$  This string (or contour) labelling of
the operators leads to an important feature of the Schwinger-Dyson
operator in gauge theories: It is {\it second-order} in
functional derivatives with respect to sources for single Wilson loop
operators.  This  is  crucial in
interpreting the Schwinger-Dyson operator as the Wheeler-DeWitt
operator.  Let us consider this in some
detail: The point is that the term
$\delta^{2}\cO_{C}/\delta A^{2}$ in equation~(\ref{gaugesd})\ 
for Wilson loops gives 
\begin{eqnarray}
\int \dd x
\sum_{\mu,a}&&{\delta^{2}\over\delta A_{\mu a}(x)^{2}} {\rm Tr~P}\ee^{i\oint 
A}
= -\sum_{a}\oint \dd s_{1}\dd s_{2} \left({\dot x(s_{1})}\cdot{\dot 
x(s_{2})}\right) \times 
\nonumber 
\\ 
&&{\rm Tr~P}\left( 
\ee^{\int_{x(s_{1})}^{x(s_{2})} A} T_{a}\ee^{\int_{x(s_{2})}^{x(s_{1})} 
A}T_{a}\right) \delta(x(s_{1})-x(s_{2})).
\end{eqnarray}
Using the identity
\begin{equation}
	\sum_{a}\Tr \left(T_{a}XT_{a}Y\right)  =
N\cdot \Tr X\cdot\Tr Y
\end{equation}
valid for U($N$) matrices, with $T_{a}$ a basis for
normalized Hermitian matrices, this  splits a
Wilson loop into two Wilson loops.
Thus, the Schwinger-Dyson equations in a gauge theory are
schematically
\begin{eqnarray}
H_{\ssc\rm SD}\ee^{W[J]}\equiv
	&&\sum_{C} \bigg[J^{C}\bigg(
	\sum_{C',C'':(C'C'')=C}\dj{C'}\dj{C''} \nonumber\\
	&&\qquad - {1\over\lambda} \dj{\hat C}\bigg)
	+ {1\over N^2}\sum_{C'}J^{C}J^{C'}\dj{(CC')}\bigg] \ee^{W[J]} = 0 .
	\label{gaeq}
\end{eqnarray}
Here we have defined $(AB)$ as the contour obtained by joining
contours $A$ and $B$ when they have a point in common, and $\hat C$
is the contour obtained by joining the `infinitesimal' contour
associated with the action $S$ into the contour $C.$ We digress
briefly to make precise the notion of  an `infinitesimal' contour.

It is important to consider the regularization of these
equations (\ref{gaeq}).  Wilson loops are composite operators and
even in a finite theory such as $N=4$ supersymmetric gauge theory,
correlation functions of composite operators must be defined with
a regularization and renormalization scheme.  In particular, the action
of a regularized gauge theory can be interpreted as a sum of Wilson
loops of size equal to the regularization length scale,
as in lattice gauge theory, and it is this interpretation that is the
definition of an `infinitesimal' contour.  The Schwinger-Dyson
equations (\ref{gaeq})  must be regularized as well, but happily
a gauge invariant regularization procedure has been developed for
these\cite{halpern}.
The fact that a regularization must be introduced at this stage
implies that there will be a normalization scale implicit in the renormalized
equations. It may be that this normalization scale is related to
the string tension of type IIB string theory.

Let us compare the structure of the
SD equations for gauge theories with
the same structure for scalar field theory.
Given  a complete set of local operators $\cO_{i}$ in a scalar
field theory, we can define
the generating function
\begin{equation}
	Z[J]\equiv \ee^{W[J^{i}]} \equiv \big\langle
\ee^{\sum_{i}J^{i}\cO_{i}}\big\rangle.
\end{equation}
If the  scalar field theory  is defined by an action $S,$ we can derive
a functional differential equation satisfied by
$\ee^{W[J]}:$
\begin{equation}
	H_{\ssc\rm SD}\ee^{W(J^{i})}\equiv
	\sum_{i} \left[J^{i}\left(\dj{i''} -  \dj{\hat \iota}\right)
	+ \sum_{ k}J^{i}J^{k}\dj{(ik)}\right] \ee^{W(J^{i})} = 0
	\label{scalarsd}
\end{equation}
with $i''$ the index associated with the operator
$\delta^{2}\cO_{i}/\delta\phi^{2},$ $\hat \iota$ the index associated
with the operator $(\delta \cO_{i}/\delta\phi)(\delta S/\delta\phi),$
and $(ik)$ the index associated with the operator
$(\delta \cO_{i}/\delta\phi(x))(\delta \cO_{k}/\delta\phi(x)).$
Thus it is not necessary to  introduce a second functional derivative with
respect to sources in the SD equations for a scalar field theory.

It is a nontrivial fact that the second-order
Schwinger-Dyson equations (\ref{gaeq}) are the equilibrium conditions of
stochastic quantization in equation~(\ref{equil}).
In gauge theories, the second-order
Schwinger-Dyson equations\cite{marc} 
expressed in terms of Wilson loops are just
the so-called loop equations\cite{loop}.

\section{Relationship between SQ and AdS/CFT}

According to the AdS/CFT duality, the  partition function of the
gauge theory coupled to sources is the exponential of the supergravity
action evaluated as  a function of the
boundary values of the supergravity fields.
The duality holds at large $N$ and large $\lambda $ for slowly varying
configurations\cite{gkp,wmalda}:
\begin{equation}
\ee^{W[J]} \equiv \big\langle \ee^{\int J {\cal O}}
\big\rangle_{\scriptscriptstyle\rm CFT}=\ee^{-S_{\ssc\rm sugra}[\hat J(J)]}\ .
\label{basic}
\end{equation}
We have written the supergravity boundary values as $\hat J= \hat
J(J)$ to indicate that rescalings may be needed depending on the
dimension of the operator $\cO.$

The  right hand side of equation (\ref{basic})  can be interpreted as
the leading order term in a
path integral representation $\Psi[\hat J]$ of a wave functional of
supergravity, where
the class of metrics we integrate over has one boundary, {\it i.e.} it is
the no-boundary wave function of Hartle and Hawking. This wave functional
$\Psi[\hat J]\approx \ee^{-S_{\ssc\rm sugra}[\hat J(J)]}$
obeys the Wheeler-DeWitt equation
\begin{equation}
H_{\ssc\rm WD} \Psi[\hat J] \approx 0
\label{wd}
\end{equation}
where the Wheeler-DeWitt operator $H_{\ssc\rm WD}$ is just the
supergravity Hamiltonian, the sum of the gravity and matter 
Hamiltonians.

Before continuing let us consider exactly how one gets to the
supergravity limit from the gauge theory. The generating function
$W[J]$ of the
gauge theory
is a function of all the possible sources. We separate these into
sources for supergravity excitations  $J_{\rm sugra}$  and sources for
the rest  which we will call string sources  $J_{\rm string}.$  In order
to obtain the effective supergravity description one does not simply
set $J_{\rm string}=0. $  The correct
prescription is to solve for $J_{\rm string}=J_{\rm string}(J_{\rm
sugra}),$ in other words to solve for massive backgrounds in terms of
slowly varying massless backgrounds.
The requirement that there is no production of string excitations
in the effective supergravity theory
in any low-energy supergravity scattering, translated into
conditions in terms of the gauge theory operators that couple
to either supergravity fields or to string fields, is
\begin{equation}
\big\langle
{\cal O}_{\rm sugra}^{1} \ldots {\cal O}_{\rm sugra}^{n}{\cal O}_{\rm
string}\big\rangle =0.
\end{equation}
In terms of $W[J]$ this is
\begin{equation}
\frac{\delta W[J]}{\delta J_{\rm string}}=0 ,
\end{equation}
which is the usual procedure of solving the equations of motion of
massive backgrounds as functions of massless backgrounds.
This in fact is exactly how the supergravity approximation is supposed
to arise from a string field theory. The nonlinear terms in the
Einstein action arise only after integrating out the massive string modes.

{}From the basic identification of the AdS/CFT duality equation
(\ref{basic}), notice that
if we find a solution of either equation (\ref{gaeq})\
or equation (\ref{wd})\ (with appropriate
boundary conditions), we solve the whole theory.
Both equations depend only on boundary values of bulk fields, and
both equations are second-order in functional derivatives.  We
emphasize again that the second-order form of the SD operator is a
consequence of the natural operators in the gauge theory being Wilson
loops.  An identification of the SD operator with the
WD operator in the supergravity  limit is therefore indicated.

Beyond the supergravity limit, the supergravity Hamiltonian (the WD operator)
must be replaced by the complete string field theory Hamiltonian.  Thus the
SD operator of the gauge theory, written in terms of its action on
Wilson loops, should be identified with the string field theory Hamiltonian.
Indeed, the structure of the SD operator that follows from its derivation
from the FP Hamiltonian has exactly the appropriate form to be a string field
theory Hamiltonian since the FP Hamiltonian takes the form
\begin{equation}
H_{\ssc\rm FP}\sim {1\over\lambda} \left[\hbox{Diffusion +
Tadpole}\right] + \left[\hbox{Loop splitting}\right]
+ {1\over N^{2}} \left[\hbox{Loop
joining}\right].
\end{equation}
The identification of the
WD operator with the SD operator suggests a relation between the
radial coordinate in AdS and the stochastic time coordinate in
stochastic quantization\cite{lp}. For example,  let
us see how the computation of a Wilson loop is viewed in stochastic
quantization.
We start with
\begin{equation}
\laa0|\ee^{-\tau H_{\ssc\rm FP}}W(C)|0\raa .
\label{wils}
\end{equation}
For a smooth non-intersecting loop,
the Fokker-Planck operator deforms the loop.  In equation~(\ref{wils}), we are
calculating the amplitude for the
loop to disappear into the vacuum. In the large 't Hooft
coupling limit ($\lambda\uparrow\infty$)
the most economical way is to continuously deform the
loop to zero size where it is annihilated by the tadpole operator.
If we write
eq.~(\ref{wils}) as a functional integral,  the path (in loop space) that the
loop takes to approach zero should then be a saddle point of the action
for large $\lambda$.
This is exactly
like the picture of the Feynman graphs at the end of  section 2, and
is also how the computation proceeds in the AdS/CFT correpondence
where the loop's path in loop space is computed using the minimal area in the
AdS background\cite{maldaloop}. Thus evolution in the stochastic time direction
is like evolving from the boundary of AdS inwards. As such the
evolution operators on both sides should be identified up to the sign
resulting from the direction of evolution, inwards (FP) or
outwards (WD).
As the AdS side correpond to looking at sources for
the Wilson loops, it is the SD operator that becomes the operator
that gives  the classical string equations in the
background.  Of
course, loops are deformed in exactly the same manner for either the Wilson
loop or its source.

We hasten to add that the above description does not imply directly
that evolution inwards from the boundary in anti-de Sitter space is
directly related to the stochastic evolution.  We must keep in mind 
that the gauge theory is recovered in the limit $\tau \uparrow\infty.$
As will become clear in the next section, even taking $\tau \uparrow\infty$
leaves a set of correlation functions that have operators separated in 
stochastic time.

\section{Bulk operators}

In semiclassical gravity there is a
a notion of approximate locality in spacetime. We should be
able to calculate properties of processes
that are approximately localized in the bulk.
We would like to construct a set of operators in the gauge theory
that reflects this property. It has been suggested that the radial
direction corresponds to
a cut-off scale or renormalization scale, but what we need is a set
of operators that can be defined
at any point in spacetime. These will then give correlation
functions at any $n$
distinct points in the bulk.
Let us label by ${\cal O}_{i}(0)$  the set of gauge theory operators
whose correlation functions are given by varying the string theory wave
functional with respect
to the boundary value of the string fields.  In the supergravity limit
these are just the local
gauge invariant operators. While this set of operators is a complete
set in the gauge theory,
its interpretation in the gravity theory seems to confine it to a
hypersurface
in  spacetime. The string theory is holographic, {\it i.e.}
all its observables  are those of the gauge theory.
In the low-energy semiclassical limit, we should still be
able to find observables of the gauge theory that can
be interpreted
as the observables of a theory in one dimension higher.

Using  the  understanding we have developed in previous sections of how
the bulk arises in the gauge theory viewed via stochastic quantization,
it is easy to see that there is a  set of operators with
the appropriate evolution in the radial direction.
As we have discussed, the SD operator, in the semiclassical limit,
is the generator of translations in the radial direction.
The SD operator does not act directly on the gauge theory
operators ${\cal O}_{i}. $ From equations~(\ref{fpsd}) and
(\ref{scf}),
we see that an appropriate
operator can be defined as
\begin{equation}
{\cal O}(t)\equiv \ee^{-tH_{\ssc\rm FP}}{\cal O}(0)\ee^{t H_{\ssc\rm FP}}
\label{dco}
\end{equation}

The right hand side of equation~(\ref{dco}) is an operator in the
original
gauge theory (although a rather unusual one).
We claim that gauge theory correlation functions with
operators as
in equation~(\ref{dco})   convey the information of the bulk theory.
Thus we will call these bulk operators. We would like to understand the
properties of these operators and the connection between the parameters $t_i$
and the physical radial distance in the AdS.

Let us look at the equation of motion of the expectation value
of a local bulk operator in the gauge theory
\begin{equation}
f[J]\equiv\langle{\cal O}(x,t) \ee^{\int J {\cal O}}\rangle ,
\end{equation}
with $\langle\ldots\rangle$ the gauge theory vacuum expectation value, as
before.
The equation of motion is then
\begin{equation}
\frac{\dd}{\dd t}f[J]=- \langle[H_{\ssc\rm FP},{\cal O}(x,t)]\ee^{\int{\cal
O}J}\rangle .
\label{eq1}
\end{equation}
Let us define $\Pi_{J}(x)\equiv {\delta}/{\delta J(x)}$, then the
above equation can be recast in terms of sources to be
\begin{equation}
\frac{\dd}{\dd \tilde{t}} \Pi_{J}(x,\tilde{t}) \langle\ee^{J{\cal O}}\rangle
=-[H_{\ssc\rm SD},\Pi_{J}(x,\tilde{t})]\langle\ee^{\int
J{\cal O}}\rangle
\label{eq2}
\end{equation}
where we have defined $\Pi_{J}(x,\tilde{t})=\ee^{-\tilde{t}H_{\ssc\rm SD}} 
\Pi_{J}(x)
\ee^{\tilde{t}H_{\ssc\rm SD}}$, and $\tilde{t}=-t$. 
The minus sign just reflects that the natural parametrization 
from the gravity point of view is 
toward the boundary of AdS while from the gauge theory it is
inwards.
Together with equation (\ref{eq2}) that we will rewrite as  
\begin{equation}
\frac{\dd}{\dd \tilde{t}} \Pi_{J}(x,\tilde{t}) 
=-[H_{\ssc\rm SD},\Pi_{J}(x,\tilde{t})]
\label{eqom}
\end{equation}
one also has the equation 
\begin{equation}
H_{\ssc\rm SD}=0.
\label{wdeq}
\end{equation}
Thus we see that if we identify the SD operator in the 
gravity limit
(which in the gauge theory was described in section 3) with the WD
operator, then
equations (\ref{eqom}, \ref{wdeq}) is just the classical equation of 
motion of the
gravity theory. What we are missing is the four-dimensional 
diffeomorphism constraint,
but it follows just from the fact that  $J(x)$ is a parameter in the  
gauge theory integral.
So our bulk operators defined above obey the classical gravity
equation
of motion.  Conversely,
since the SD operator in the semiclassical limit is hard to
compute, we can start instead with the WD operator. The WD
operator written in terms of supergravity fields and momenta (sources
$J$ and $\Pi_{J}$) can be
rewritten as an operator in the gauge theory variables (${\cal O}$ and
$ {\delta}/{\delta {\cal O}}$). If we now use this operator
(which is easy to compute starting from the supergravity) to
construct the
gauge theory bulk operators as in equation (\ref{dco}), we are
guaranteed to
obtain operators which obey the supergravity equations of motion.

Let us examine in more detail how the semiclassical limit will emerge.
The label $t$ which we have been using  is
not the physical radial distance, for any semiclassical space time. The 
semiclassical
limit is the one in which we first solve for the classical part of the metric
(in the AdS case this is just the conformal part of the metric). The value
of the metric plays the role of the physical radial distance. This then can be
substituted
back into the equation for the fluctuations about the background in order to get
the classical evolution of the fields on a fixed spacetime. The parameter
$t$ that we have introduced is similar to the foliation
label one introduces when quantizing gravity, so the wave function
is independent of this parameter. Only after we have solved the metric
equations and defined physical time can we connect the parameter $t$ to the
radial direction of the AdS. This is evident for instance
from the property of translation in $t$ of the bulk correlation functions.

Once this is done however the bulk correlation functions will be identified
with local bulk information at various bulk positions. Thus any properties
of the correlation functions as a function of the $t_i$ are translated
into properties of
supergravity correlation functions in the bulk. An example of this
was just given above in terms of the equations of motion.
However our definition of the bulk operators is
valid beyond the classical limit.

In the AdS/CFT connection there is some evidence for a
relationship between the RG flow and propagation in the radial direction.
A relation between stochastic time and the renormalization 
group is evident from the identification of finite stochastic 
time $\tau$ amplitudes  (equation~(\ref{cor1})) as infrared
regulated amplitudes\cite{ikspecial} 
(see {\it e.g.} equation~(\ref{finiteir})).
To further our understanding 
let us look at the RG equation for the $d+1$ dimensional
stochastic field theory $S_{\scriptstyle\rm stoc}$ defined in section 2.
Until now we have been relatively na\"\i ve, ignoring the issue of
renormalization.
Stochastic Ward identities restrict the possible renormalization
functions
of the stochastic theory to be those of the original theory plus
renormalization
of the stochastic time scale\cite{zinn}, 
which we have labelled  $\Omega$ in the
Langevin equation (\ref{lan}).

Thus if we introduce a renormalization scale $\mu$, we have\cite{zinn}
\begin{equation}
\mu \frac{\partial}{\partial \mu} \log \Omega =\eta_{\omega}(g).
\end{equation}
The renormalization group equation for the Green functions of the
stochastic theory is then
\begin{equation}
\left[RG_{\ssc d}+\eta_{\omega}(g)\Omega \frac{\partial}{\partial
\Omega}\right]\Gamma^{n}(x_{i},t_{i};\mu,\Omega)=0,
\label{brg}
\end{equation}
where $RG_{\ssc d}$ stands for the renormalization group operator in
the $d$ dimensional gauge theory.
For a conformal field theory (for which the couplings do not run) one
has a solution
($\eta_{\omega}(g)$ is just a number)
\begin{equation}
\frac{\Omega}{\Omega_{0}}=\left(\frac{\mu}{\mu_{0}}\right)^{\eta_{\omega}(g)},
\label{eta}
\end{equation}
{\it i.e.} if $\mu$ is rescaled so is $\Omega.$  Since
$\Omega$ is just the scale for the parameter $t$ that is related
to the radial direction in the semiclassical limit as explained above,
this establishes a connection between the radial direction and the
RG flow.   

Attempts to directly link supergravity equations of motion to
the renormalization group can be found in \cite{bvv}; see also 
\cite{kost}.

Another  property of objects in the bulk in the AdS/CFT relationship
is that objects in the interior look nonlocal from the perspective
of the boundary theory.
To see how this arises from the construction of the bulk operators
let us look at the free scalar field $\phi$ in four dimensions.
The simplest example is to look at
the two point function
of the operators correponding to those in equation (\ref{dco}):
\begin{equation}
\langle\phi(k,t)\phi(-k,0)\rangle=\frac{1}{k^2}\ee^{-k^2t},
\label{nonloc}
\end{equation}
or in the space representation,
\begin{equation}
\langle\phi(x,t)\phi(x',0)\rangle
\sim\frac{1}{|x-x'|^2}\left[1-\ee^{-|x-x'|^2/4t}\right].
\end{equation}
We see that compared with a two point function of two operators at the same
$t$
the two point function of operators at differing values of $t$ has an
effective UV cutoff. While this example was in the free field case, this
property still holds when interactions are turned on.
This is exactly in acord with
what one expects in the AdS/CFT, but here we see the
details of how it works.

This property however should not be confused with the locality of these
operators in the bulk. From the supergravity perspective this is
a confusing point. If we assume that an operator in the bulk is nonlocal
with some nonlocal scale related to the position in the bulk, then
how is it that the supergravity still looks local on scales which are
smaller than this nonlocal scale? Indeed in the AdS/CFT using the metric
\begin{equation}
\dd s^2=l_{s}^{2}\left[\frac{U^2}{\sqrt 2\lambda}(\dd x^{\nu}\dd x_{\nu})+
\frac{\sqrt{2\lambda}}{U^2}\dd U^2+\sqrt{2\lambda} \dd 
\Omega_{5}^{2}\right],
\end{equation}
objects at coordinate $U$ are assumed to have a nonlocal 
scale\cite{pp}\ 
$\frac{\sqrt\lambda}{U}$,
but supergravity should be valid down  to the string scale which is
$\frac{\root 4 \of \lambda}{U}$ which is much less than the nonlocal scale
for large 't Hooft coupling $\lambda$.
Using the bulk operators this is easy to understand.
If we look at some correlation function
\begin{equation}
\langle{{\cal
O}_{3}(x_3,t_3)\cal O}_{2}(x_2,t_2){\cal O}_{1}(x_1,0)\rangle ,
\end{equation}
it has no singularities as the $x_i$ coincide, but if $t_2 \sim t_3 \neq 0$
the correlation function will have singularities when $x_{2}$ and $x_{3}$
are close together ({\it i.e.} ${\cal O}_{2}$ and ${\cal O}_{3}$
behave as local operators with respect to each other), while with
respect to ${\cal O}_{1}$ they behave as nonlocal operators.

If we  assemble some local excitations in the bulk all at  some
$t_{i}>t_0,$ and probe them with
local probes in the gauge theory that are at $t=0,$ equation (\ref{nonloc})
tells us that  from the point of view of the local operators of
the gauge theory, the operators in the bulk are only correlated with
the low momentum modes of operators at $t=0.$
This is as it should because  a bounded  region inside
the bulk can only support a finite entropy.  This is
of course related to holography. If we take some UV cutoff for the
gauge theory, then there are a finite number of degrees of freedom
(for finite $N$), and of course a finite number of independent
approximately local operators in the gauge theory.
Using the bulk
operators we can construct matter distributions inside the anti-de
Sitter space,  and try to exceed
the Bekenstein bound. All the bulk operators can be written as
sums of the approximately local operators in the gauge theory, and
therefore the number of independent bulk operators cannot exceed the
number of degrees of freedom of the original gauge 
theory. 

Another benefit of our construction of bulk operators
is the understanding of cluster
decomposition in the bulk: Why are correlation functions of operators at
very different radial positions $(t)$ and similar transverse positions
$(x)$ suppressed?  For different transverse positions this is  because of the
locality of the original theory,
but $t$ is not a coordinate in the original space.
However  the evolution in the   $t$ direction is generated by the
Fokker-Planck Hamiltonian.
As we have seen in section 2 regarding the Langevin equation (\ref{lan}),
the Fokker-Planck Hamiltonian generates
an approach to equilibrium driven by a noise.  Perturbing the state
by the insertion of an operator, and looking at its approach back to
equilibrium
is just what is computed in correlation functions with operator
insertions at different times. General
properties of the approach to equilibrium can then be used to show that
correlation functions of operators at different times fall off with
the time difference. In fact for a theory with a mass gap they
decay exponentially and with no mass gap they decay with a power law.

\section{Conclusions and further discussion}

In this paper we have argued that the gauge invariant
Schwinger-Dyson operator has a natural
dual in the dual supergravity theory as the supergravity Hamiltonian conjugate
to the radial foliation  of the AdS. It will become
the string field theory Hamiltonian away from the semi-classical limit.
Using this we have shown that there is a natural (though far from simple)
class of observables that mimics the existence of an extra large
dimension. Properties of these observables were shown to coincide with
general expectations, but the construction enables one to understand
the connection between the gauge theory and the supergravity in a deeper way.

One interesting question that
can be addressed with this identification is the issue of Minkowski
holography.  In our formalism, this question turns into: What gauge
theory action has a Schwinger-Dyson operator that is equal to the
Wheeler-DeWitt operator for a vanishing cosmological constant?

\section{Acknowledgements}
We would like to thank A. Polyakov and H. Verlinde for discussions.
G.L. would like to thank W. Fischler, I. Klebanov,  
S. Shenker, L. Susskind  and E. Verlinde for  helpful conversations.
V.P. is grateful to B. Durand and L. Durand for their hospitality at
the University of Wisconsin, Madison.
This work was supported in part by NSF grant PHY98-02484.

\def\np#1#2#3{Nucl. Phys. B#1,  #3 (#2)}
\def\prd#1#2#3{Phys. Rev. D#1, #3 (#2)}
\def\prl#1#2#3{Phys. Rev. Lett. #1, #3 (#2)}
\def\pl#1#2#3{Phys. Lett. B#1, #3 (#2)}


\begin{thebibliography}{99}

\bibitem{malda} J. Maldacena, Adv. Theor. Math. Phys. 2, 231 (1998) 

\bibitem{gkp} S. Gubser, I. Klebanov and A. Polyakov, \pl{428}{1998}{105}

\bibitem{wmalda} E. Witten, Adv. Theor. Math. Phys. 2, 253 (1998) 


\bibitem{thooft} G. 't Hooft, \np{72}{1974}{461}

\bibitem{caubh} D. Kabat and G. Lifschytz, JHEP 9812, 002 (1998); JHEP 
9905, 005 (1999); S. Das, JHEP 
9902, 012 (1999); G. Horowitz and N. Itzhaki, JHEP 9902, 010 (1999)

\bibitem{bdhm} 
T. Banks, M. Douglas, G. Horowitz and E. Martinec, {\sl
AdS dynamics from conformal field theory}, hep-th/9808016;
V. Balasubramanian, P. Kraus, A. Lawrence and S. Trivedi, 
\prd{59}{1999}{104021} 

\bibitem{th} G. 't Hooft, in {\it Salamfestschrift: a collection of
talks}, World Scientific Series in 20th century physics, v. 4, eds. A. Ali
et al. (World Sci., 1993) and in Proc. Symp. {\it The Oskar Klein
Centenary}, ed. U. Lindstr\"om (World Sci., 1995); L. Susskind, J. Math.
Phys. 36, 6377 (1995)

\bibitem{witsus} L. Susskind and E. Witten, {\sl The holographic bound 
in anti-de Sitter space}, hep-th/9805114

\bibitem{ik} N. Ishibashi and H. Kawai, \pl{314}{1993}{190};
M. Fukuma, N. Ishibashi, H. Kawai and M. Ninomiya,
\np{427}{1994}{139}; M. Ikehara, N. Ishibashi, H. Kawai, T. Mogami, R.
Nakayama and N. Sasakura, \prd{50}{1994}{7467};
N. Ishibashi and H. Kawai, \pl{322}{1994}{67};
\pl{352}{1995}{75}


\bibitem{jevic} A. Jevicki and J. Rodrigues, \np{421}{1994}{278}; see
also S. Das and A. Jevicki, Mod. Phys. Lett. A5, 1639 (1990);
G. Moore, N. Seiberg and M. Staudacher, \np{362}{1991}{665}

\bibitem{pw} G. Parisi and Y.-S. Wu, Sci. Sin. 24, 484 (1981)

\bibitem{marc} G. Marchesini, Nucl. Phys. B191, 214 (1981); B239,
135 (1984)

\bibitem{v1} V. Periwal, {\sl String field theory Hamiltonians from
Yang-Mills theories:toy model of Polyakov duality}, 
hep-th/9906052, to appear in Phys. Rev.

\bibitem{pol} A. Polyakov, Nucl. Phys. Proc. Supp. 68, 1 (1998);
Int. J. Mod. Phys. A14, 645 (1999)

\bibitem{loop} G. De Angelis, D. De Falco and F. Guerra, Nuovo Cim.
Lett. 19, 55 (1977); F. Guerra, R. Marra and G. Immirzi, Nuovo Cim.
Lett. 23, 237 (1978); J.--L. Gervais and A. Neveu,
\pl{80}{1979}{255}; Y. Nambu, \pl{80}{1979}{372}; 
E. Corrigan and B. Hasslacher,
\pl{81}{1979}{181}; A. Polyakov,
\pl{82}{1979}{247}; L. Durand and E. Mendel, \pl{85}{1979}{241}
D. Foerster, Phys. Lett. 87B, 83 (1979); T. Eguchi,
Phys. Lett. 87B, 91 (1979); Yu. Makeenko
and A. Migdal, Phys. Lett. 88B, 135 (1979);
A. Jevicki and B. Sakita, Nucl. Phys. B185, 89 (1981)

\bibitem{lp} G. Lifschytz and V. Periwal, {\sl Dynamical truncation of the
string spectrum at finite $N$}, hep-th/9909152 

\bibitem{others} S. Hirano, {\sl Exact renormalization group and loop 
equation}, hep-th/9910256; M. Li, {\sl A note on relation between 
holographic rg equation and Polchinski's rg equation}, hep-th/0001193;
C. van de Bruck, {\sl On gravity, holography and the quantum}, 
gr-qc/0001048 


\bibitem{stoc} P. H. Damgaard and H. Huffel, 
Physics Reports {  152}, 227 (1987); J. Zinn-Justin {\sl Quantum field
theory and critical phenomena}, Oxford University Press (1996) 

\bibitem{ikspecial} M. Ikehara, N. Ishibashi, H. Kawai, T. Mogami, R. 
Nakayama and N. Sasakura, Prog. Theor. Phys. Suppl. 118, 241 (1995)

\bibitem{halpern} M.B. Halpern, Prog. Theor. Phys. Suppl. 111, 163 (1993)

\bibitem{maldaloop} S.-J. Rey 
and J. Yee, {\sl Macroscopic strings as heavy quarks in large $N$ 
gauge theroy and anti-de Sitter supergravity}, hep-th/9803001;
J.M. Maldacena, \prl{80}{1998}{4859} 

\bibitem{zinn} J. Zinn-Justin, \np{275}{1986}{135};
Prog. Theor. Phys. Suppl. 111, 185 (1993)

\bibitem{bvv} J. de Boer, E. Verlinde and H. Verlinde, {\sl On the
holographic renormalization group}, hep-th/9912012

\bibitem{kost} S. de Haro, K. Skenderis and S.N. Solodukhin,
{\sl Holographic reconstruction of spacetime and renormalization in 
the AdS/CFT correspondence}, hep-th/0002230


\bibitem{pp} A. Peet and J. Polchinski, \prd{59}{1999}{065011}





















\end{thebibliography}
\end{document}